# Evaluation of the Bychkov- Rashba Field from the Spin Resonance of Electrons in a Si Quantum Well

Z. Wilamowski[1,2], W. Jantsch[1]

[1]*Institut für Halbleiterphysik, Johannes Kepler Universität, A-4040 Linz, Austria*
[2]*Institute of Physics, Polish Academy of Sciences, Al Lotnikow 32/46, PL 0668 Warsaw, Poland*

From spin resonance of two-dimensional (2D) conduction electrons in a modulation doped SiGe/Si/SiGe quantum well structure we find a 2D anisotropy of both the line broadening (dephasing time) and the g-factor. We show that these can be explained consistently in terms of the Bychkov-Rashba (BR) field which here is the dominant coupling between electron motion and spin. We obtain a BR parameter of $\alpha = 1.1 \cdot 10^{-12}$ eV·cm - three orders of magnitude smaller as compared to III-V wells. Extrapolating for low electron concentrations we obtain a g-factor of the Si conduction band of $2.00073 \pm 0.00010$.

The present attempts to construct a spin based electronic device, where spin switches or spin processors (calculators) are controlled by classical electronics, require a detailed understanding of spin properties and their relation to the electronic properties, or, in other words, a detailed knowledge of spin-orbit coupling.

Conduction electrons are natural candidates for such devices. Especially, in low dimensional "nano"-structures, the conduction electrons can be easily manipulated by applied voltages or by illumination with light. Moreover, as it has been shown already for two[1,2,3] -dimensional (2D) quantum wells, the electrical conductivity depends on the spin polarization. Electrical measurements can be used thus to read the spin state. In this paper we investigate the effect of concentration of a 2D electron gas (2DEG) on the spin properties. We look for the microscopic mechanisms which rule the spin properties of the 2D electrons. We conclude that the so called Bychkov-Rashba[4,5] (BR) term affects the basic spin properties: the g-shift, its anisotropy, and the spin relaxation. We demonstrate and make use of the fact that conduction electron spin resonance (CESR) provides a very effective and sensitive tool for the evaluation of details of spin orbit coupling and in particular, to evaluate the BR field. This CESR method is by orders of magnitude more sensitive as compared to methods based on: (i) beating effects in magneto oscillations[6,7], (ii) Raman scattering[8] or (iii) the investigation of weak localization[9,10].which were applied to III-V semiconductors where the zero field spin splitting is much bigger.

As pointed out by Rashba and Bychkov[4,5], for axial systems with broken mirror symmetry an additional spin dependent term is needed in the Hamiltonian, $\mathcal{H}$, in addition to the Zeeman term. This BR term can be described by the vector product of the **k**-vector and the spin operator, **σ**:

$$\mathcal{H} = g_0 \mu_B \mathbf{H}_{ex} \boldsymbol{\sigma} + \alpha (\mathbf{k} \times \boldsymbol{\sigma}) \hat{z} \qquad (1)$$

Here, $g_0$ is the g-factor, $\mu_B$ the Bohr magneton, $\mathbf{H}_{ex}$ the applied magnetic field, $\hat{z}$ is the direction of broken symmetry and the parameter $\alpha$ depends on the spin-orbit coupling and on the details of the crystal symmetry and, in our case, a quantum well structure, which lacks mirror symmetry[11,12] mostly because of one-sided modulation doping. The role of the BR term has been extensively studied in the context of weak localization[10] and also of the metal-to-insulator transition in the 2DEG.[13,14] The BR term has recently received also much attention as a possible mechanism to manipulate spins as required for spin transistors[15] or quantum computing[16] in materials with large α. On the other hand, since the BR term leads to an additional channel for spin lattice relaxation, materials with small α are preferable if long spin life times are needed.

The BR term causes spin splitting. Because of that, it can be treated as an effective magnetic field acting on the electron. This BR field does not vanish at zero field, it lies in the 2D plane and it is perpendicular to **k**:

$$\mathbf{H}_{BR} = \frac{\alpha \mathbf{k} \times \hat{z}}{g \mu_B} = \frac{\alpha}{g \mu_B} \sqrt{\frac{2\pi n_s}{g_v}} \frac{\mathbf{k} \times \hat{z}}{k} \qquad (2)$$

Here $g_v$ stands for valley degeneracy factor with $g_v = 2$ for our strained Si well grown on [100] substrate.

In our standard ESR experiment we are able to measure CESR of the high mobility 2DEG in modulation doped Si/SiGe structures and simultaneously cyclotron resonance (CR).[3,17] The latter allows us to determine the carrier density, $n_s$, and momentum relaxation rate, $\tau_k$, *in situ*.[17] Making use of the persistent photo-conductivity in our structures[17] we follow the dependence of CESR parameters on $k_F$, (determined by the electron concentration), the direction of applied magnetic field, momentum relaxation rate $1/\tau_k$ and temperature. Experimental details and sample structure were described in detail in Refs. 3 and 17.

Experimental results for the conduction electronic g-factor in Si quantum wells are given in Fig. 1. The g-factor is practically temperature independent in the whole range of temperatures investigated – from 2 to 50 K. The g-factor for a magnetic field perpendicular to the layer ($\Theta=0$) is bigger than for in-plane orientation. The anisotropy of the g-factor, $\Delta g = g(0^o) - g(90^o)$, is shown in Fig. 1a as a function of $n_s$ and in Fig. 1b its mean value, defined as: $\langle g \rangle = [g(0^o) + 2g(90^o)]/3$ is plotted. Both quantities vary linearly with $n_s$.

The CESR of 2D electrons in Si/SiGe structures is characterized by a very narrow line width, $\Delta H$, which depends on the direction of applied magnetic field. For in-plane orientation the line width is typically a few tenths of a Gauss, but for perpendicular orientation the line width becomes smaller by an order of magnitude. The narrowest line width we observed is 30 mG (see Fig. 2a), about two orders of magnitude narrower[17] than that of typical ESR lines of defects in Si. The line width for in-plane orientation is plotted in Fig. 2a as a function of $n_s$. We find (see below) that this line width, expressed usually by the transverse relaxation rate, $\Delta_2$, exceeds the one caused by finite spin life time, $\Delta_1 = 1/\tau_s < \Delta_2$. For metals, the opposite is true since the spin life time is limited by momentum scattering. In our samples this is apparently not the case because of the high mobility.

In order test the mechanism of spin relaxation we investigate also the momentum scattering rate which manifests itself in the CR line width. Fig. 2b shows the CR line width as a function of $n_s$ as obtained from three samples with different doping concentrations.[3,17] For low $n_s$, close to the metal-insulator transition, the CR line width shows a tendency to diverge as the potential fluctuations diverge due to breakdown of screening.[3] For the Elliott-Yaffet mechanism the spin-flip probability is expected to be proportional to the momentum scattering rate.[10] Taking the dependencies of Figs. 2a and b, we observe an inverse proportionality indicating rather a Dyakonov-Perrel-like mechanism[10] where momentum scat-tering causes motional narrowing.

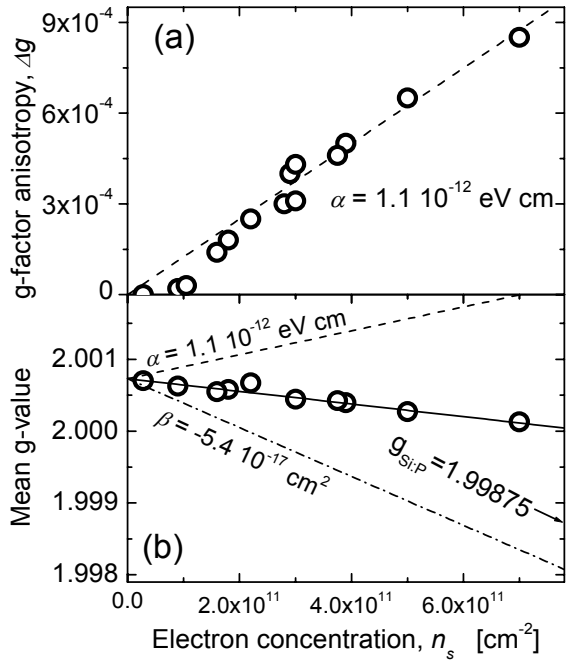

**Fig. 1a:** Measured g-factor anisotropy (circles), $\Delta g = g(0^o) - g(90^o)$, vs. electron concentration. Dashed line: fit resulting in a BR parameter of $\alpha=1.1\cdot 10^{-12}$ eV cm.
**Fig. 1b:** mean g-value. Dashed line: Rashba contribution obtained with the same $\alpha$ parameter as in (a), dash-dotted line: effect of non-parabolicity with $\beta$ adjusted to achieve agreement of the combination of the two effects with experiment (full line).

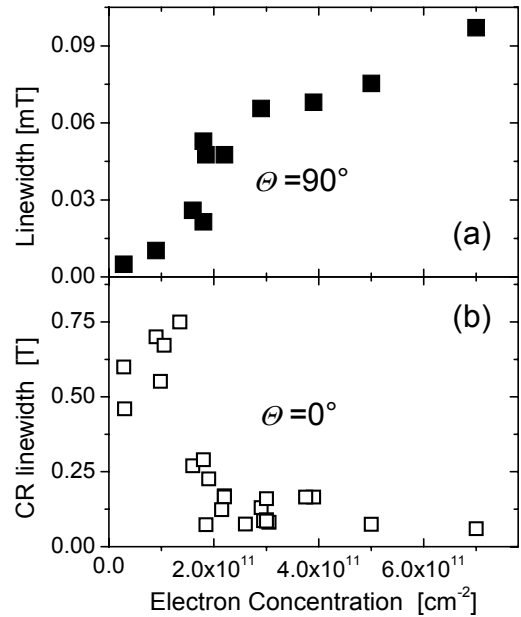

**Fig. 2:** (a) CESR line width for in-plane magnetic field and (b) cyclotron resonance linewidth for perpendicular magnetic field.

For Fig. 1, the g-factor is evaluated as usual from the externally applied field at which resonance occurs, $\mathbf{H}_{ex}$, and the microwave frequency. In the presence of the BR field, $\mathbf{H}_{BR}$, however, the resonance condition takes the form: $\hbar\omega = g_o \mu_B |\mathbf{H}_{eff}|$, where $g_o$ is a native electron g-factor and the effective magnetic field is $\mathbf{H}_{eff} = \mathbf{H}_{ex} + \mathbf{H}_{BR}$. When the external field, $\mathbf{H}_{ex}$, is applied at an angle $\Theta$, and $\mathbf{H}_{BR}$ is distributed in the surface of the layer then the value of $\mathbf{H}_{eff}$ is distributed around its mean value, $<\mathbf{H}_{eff}>$. For $\mathbf{H}_{BR} \ll \mathbf{H}_{ex}$, the variance of the distribution of the effective local field, which contributes to the line width broadening, is:

$$\delta H_{BR}^2 = \frac{1}{2} H_{BR}^2 \sin^2 \Theta \quad . \quad (3)$$

The dependence of the measured g-factor on the electron concentration, $n_s$, and on the direction of the external magnetic field is described by:

$$g(n_s, \Theta) = g_o \frac{\langle H_{eff} \rangle}{H_{ex}} \cong g_o \left(1 + \frac{H_{BR}^2}{4 H_{ex}^2}\left(1 + \cos^2 \Theta\right)\right), \quad (4)$$

where $H_{BR}$ is given by Eq. 2. The right hand expression was evaluated assuming a degenerate 2DEG, i.e., the observed g-factor corresponds to the g-value at $k_F$ only. Eq.(4) shows that the BR coupling leads to a g-factor anisotropy. From the observed g-factor anisotropy we are thus able to extract $H_{BR}$ without any adjustable parameters and the results are given in Fig. 3. The solid line in Fig. 3 is obtained from Eq. 2 using $\alpha = 1 \cdot 10^{12}$ eV·cm as a fitting parameter. The BR coefficient $\alpha$ turns out to be independent of $n_s$ as it can be seen from the linear increase of the g-factor anisotropy with electron concentration (s. Fig. 1a).

For $n_s = 4 \cdot 10^{11}$ cm$^{-2}$, the value of $\alpha$ obtained gives a Rashba field of $H_R = 100$ G which yields a zero-field splitting of a few µeV, about three orders of magnitude smaller than for InAs wells.[6]

The dashed line in Fig. 1b represents data for $<g>$ obtained from Eq. 4 using the same value of $\alpha$. Considering only the BR effect, obviously an increase of $<g>$ is expected in contrast to experiment. Here we have to correct for the isotropic g-shift, $g_0 = g_{00}(1+\beta k^2)$, that results from non-parabolicity (dash-dotted line in Fig.2b). We can explain the observed dependence of $<g>$ on $n_s$ thus in terms of the sum of the Rashba- and the non-parabolicity shifts using a value of $\beta = -5.4 \cdot 10^{-17}$ cm$^2$. The value of $<g>$ extrapolated for zero concentration, $g_{00} = 2.00073 \pm 0.00010$, (conduction band edge) is considerably bigger than that given by Feher[18] for heavily doped Si:P bulk material ($g_{Si} = 1.99875 \pm 0.00010$). This difference is again a consequence of the non-parabolicity and can be explained with the same β value.

The BR field is perpendicular to the **k**-vector of the electron under consideration. Since the Fermi vector, $\mathbf{k}_F$, may have any direction in the 2D plane, the BR field has also any direction in plane. Since the directions of $\mathbf{H}_{BR}$, are randomly spread within the 2D layer the resonance of the electrons occurs at different external fields. As a result, the line becomes broadened by the BR spin-orbit coupling. Neglecting both momentum scattering and cyclotron motion, one could expect that the line width broadening due to the BR effect would be described by the variance of the magnetic field seen by the electrons as described by Eq.(3). A change of the electron momentum, which is equivalent to a change of the BR field, leads, however, to an averaging of the line width. If the characteristic frequency of modulation is much higher than the frequency corresponding to the resonance line width, then the standard formula for motional narrowing can be applied[9,10]:

$$\Delta H = \gamma \cdot \delta H_{BR}^2 \cdot \tau_k = \frac{\gamma H_{BR}^2 \sin^2 \Theta}{2} \tau_k \quad . \quad (5)$$

Here $\gamma = g\mu_B/\hbar$ is the gyromagnetic ratio and $\tau_k$ the momentum relaxation time. The observed angular dependence $\propto \sin^2\Theta$ is well described by Eq. 5 and it is a unique attribute of the discussed broadening mechanism where the fluctuating field has only an in-plane component. Taking experimental data from Fig. 2 for $\tau_k$ and the line width $\Delta H$, we can evaluate the BR field from Eq. 5. Results are also

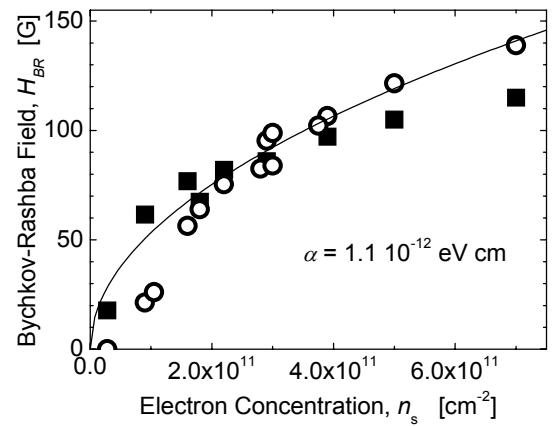

**Fig. 3:** BR field, as evaluated from the line width (■) and from the g-factor anisotropy (○) as a function of the density of a 2DEG.

3/4

given in Fig. 3 (squares). Without any further fitting parameter we obtain good agreement with the data derived from the g-factor anisotropy.

For perpendicular field ($\Theta=0$) the second moment of the distribution vanishes (s. Eq (5)) and the BR mechanism of the line broadening does not contribute anymore. For perpendicular orientation of the external magnetic field only this additional broadening is seen which is caused by the sample inhomogeneity. Some samples, especially multi-well samples and those with low mobility exhibit much larger line widths than those of Fig. 2.

Discussing the anisotropy of the g-factor we neglect the anisotropy of $g_0$, which, by symmetry, could exist. But because this type of g-factor anisotropy does not affect the line width and the values of $\alpha$ obtained from the g-factor anisotropy and from the line width are very close, we can conclude *a posteriori*, that the anisotropy of $g_0$ is negligible.

In summary, we have shown, that CESR measurements provide a very sensitive tool to evaluate the BR constant. The BR field is the origin of the observed anisotropies of the g-factor and the line width. For perpendicular field the BR line width broadening vanishes and the CESR is easiest to observe for that orientation. In Si quantum wells we observe a very weak BR field because of weak spin orbit coupling. Therefore spin relaxation is also very slow. The extremely small CESR line width (long dephasing time) and very long $T_1$ of the order of $10^{-5}$s makes Si an interesting candidate for spintronic devices.

**Acknowledgement:** We thank F. Schäffler (JKU) for generously providing samples and helpful discussions. Work supported within the KBN grant 2 P03B 007 16 in Poland and in Austria by the *Fonds zur Förderung der Wissenschaftlichen Forschung*, and ÖAD, both Vienna.